\documentstyle[psfig]{mn}
 
\def\refitem{\par\parskip 0pt\noindent\hangindent 20pt}
\def\lesssim{\mathrel{\hbox{\rlap{\hbox{\lower4pt\hbox{$\sim$}}}\hbox{$<$}}}}
\def\gtrsim{\mathrel{\hbox{\rlap{\hbox{\lower4pt\hbox{$\sim$}}}\hbox{$>$}}}}

\title[Testing the FR~I/BL Lac unifying model with HST] 
{Testing the FR~I/BL Lac unifying model with HST observations}

\author[Alessandro Capetti and Annalisa Celotti]
{Alessandro Capetti$^{1,2}$ and Annalisa Celotti$^{2}$\\
$^1$Osservatorio Astronomico di Torino, Strada Osservatorio 20,
10025 Pino Torinese, Italy\\
$^2$S.I.S.S.A., via Beirut 2--4, 34014 Trieste, Italy\\
\\}

\date{Received ***; in original form ***}       
\begin{document} 
\maketitle 

\begin{abstract}

Hubble Space Telescope (HST) observations provide a novel way of
testing unified models for FR~I radio sources and BL Lac objects.  The
detection of extended dust discs in some radio galaxies provides
information on their jet orientation. Given this, the strength of the
compact nuclear sources of FR~I and BL Lacs can be compared with model
predictions.

As a pilot project towards using HST information in testing unified
models, we selected five radio galaxies which show extended nuclear
discs in the HST images.  The relative orientation of the projected
radio-jets and of the extended nuclear discs indicates that they are
not perpendicular, as the simplest geometrical model would
suggest, but that they form an angle of $\sim 20^\circ - 40^\circ$
with the symmetry axis of the disc: a significant change of
orientation occurs between the innermost AGN structure and the
kpc-scale. Nevertheless, the discs appear to be useful indicators of
the radio source's orientation since the angles formed by the disc axis
and the jet with the line of sight differ by only $\sim 10^\circ -
20^\circ$.

At the center of each disc an unresolved nuclear source is present. We
compared its luminosity with the optical core luminosity of BL Lacs
selected for having similar host galaxy magnitude and extended radio
luminosity. The BL Lac cores are between $2 \times 10^2$ and $3 \times
10^5$ times brighter than the corresponding radio galaxies ones.

The FR~I/BL Lac core luminosity ratio shows a suggestive correlation
with the orientation of the radio galaxies with respect to the line of
sight. The behavior of this ratio is quantitatively consistent with a
scenario in which the emission in the FR~I and BL Lac is dominated by
the beamed radiation from a relativistic jet with Doppler factor $\sim
5 - 10$, thus supporting the basic features of the proposed
unification schemes.

Several observational tests, based on the method proposed here, can
strengthen our conclusions and improve the statistical significance of
the findings presented.

\end{abstract}


\section{Introduction} 

Unification schemes focus on the essence of Active Galactic Nuclei
(AGN), by separating physical properties from
geometrical/orientational effects. In fact, if anisotropic emission
occurs, intrinsically identical sources observed at different
orientations manifest different properties.

The most promising unifying model for weak radio--loud AGN identifies
low luminosity Fanaroff--Riley (1974), FR~I radio sources as the
parent population of BL Lac objects. The non--thermal continuum
emission of BL Lacs would be in fact dominated by beaming effects,
resulting from the observation of plasma moving at relativistic speed
at a small angle with respect to its direction of motion (Blandford \&
Rees 1978).  FR~I radio galaxies would then represent the mis-oriented
counterparts of BL Lacs (e.g. Urry \& Padovani 1995 for a recent
review).

Circumstantial evidence for this unification model includes the power
and morphology of the extended (supposedly unbeamed) radio emission of
BL Lacs (e.g. Antonucci \& Ulvestad 1985, Murphy et al. 1993, 
Kollgaard et al. 1992) and the properties of their host
galaxies (e.g.  Ulrich 1989; Abraham, McHardy \& Crawford 1991;
Stickel et al. 1991, 1993; Falomo et al.  1997) which are similar to
those of FR~I. Furthermore, the quantitative agreement among degrees
of beaming (angles and velocities) required by different observational
properties (e.g.  Ghisellini et al. 1993) and the comparison of number
densities and luminosity functions of the parent and beamed
populations in different bands (e.g. Padovani \& Urry 1990, 1991;
Urry, Padovani \& Stickel 1991; Celotti et al. 1993), are basically
consistent with the proposed scheme.

In this framework, we expect non-thermal emission from the jet,
although not amplified or even de-amplified, to be present also in
FR~I.  In the case of the nearest FR~I radio source Centaurus A, the
estimate of the incident continuum on highly ionized filaments
(Morganti et al. 1991, 1992) and the detection of a polarized infrared
nuclear component (Bailey et al. 1986), suggest that indeed highly
beamed emission might be present in this radio galaxy. However these
results do not appear to be conclusive, since both effects can be
accounted for without requiring non--thermal collimated emission
(Sutherland, Bicknell \& Dopita 1993, Schreier et al. 1996, Packham et
al. 1996).

Thanks to HST data, it is now possible, for the first time, to
usefully compare the FR~I nuclear emission with that observed in
BL Lacs, taking also into account their orientation. Depending on the
jet speed and direction with respect to the line of sight, the FR~I
intensity is expected to be orders of magnitude weaker than in a
corresponding BL Lac object. While at ground based resolution such a
weak nuclear non-thermal component would be totally swamped by the
stellar emission of the host galaxy, the high spatial resolution of
HST allows us to detect point sources at the center of a typical radio
galaxy as faint as V $\sim$ 25 mag.

At the same time, the HST observations have revealed the presence of
extended nuclear discs in several radio galaxies (Jaffe et al. 1993;
De Koff et al. 1995; De Juan, Colina \& Golombek 1996). These
structures have been naturally identified with the reservoir of
material which will ultimately accrete into the central black hole
(e.g. Jaffe et al. 1993). Although the precise relationship between
the symmetry axis of these discs and that of the sub-parsec scale
accretion discs is not yet established, they can represent indicators
for the orientation of the central engine.  By combining the
information on the relative intensity and orientations of the nuclear
component of the parent and beamed objects, it is possible to test its
consistency with the considered unifying scheme.

And this is the aim of this paper. As a pilot project, we selected five
radio galaxies which show extended nuclear discs in the HST images and we
then compare the intensity of the optical and radio nuclear emission of the
sample of 5 FR~I radio galaxies with those observed in the corresponding
BL Lacs, where the latter ones have been selected on the basis of
extended/unbeamed properties similar to their putative parent galaxy.

The description of the HST data for the FR~I radio galaxies is the
subject of Section 2, while in Section 3 we discuss the geometrical
disc properties of the extended nuclear discs and their relationship
with the orientation of the central sources. In Sections 4 and 5 we
give a detailed description of the comparison between radio galaxies
and BL Lacs and in Section 6 we present the results. A discussion of
the findings together with our conclusions and future prospects are
presented in Section 7.

\section{HST observations} 

We select five FR~I radio galaxies in which archival HST images
(available up to May 1997) revealed the presence of an extended
nuclear disc, namely 3C~31, 3C~264, 3C~465, NGC~4261 and NGC~7052. The
images for these objects were taken using the Wide Field and Planetary
Camera 2 (see Table~1 for a log of the observations). The pixel size
of the Planetary Camera is 0$^{\prime\prime}$.0455 and the 800
$\times$ 800 pixels cover a field of view of $36^{\prime\prime} \times
36^{\prime\prime}$. Observations were obtained in either a medium or
broad-band filter for each source. For 3C~31 and 3C~465 narrow-band
images centered on the redshifted H$\alpha$ emission are also
available.

The data were processed through the PODPS (Post Observation Data
Processing System) pipeline for bias removal and flat fielding
(Burrows et al. 1995). Individual exposures in each filter were
combined to remove cosmic rays events. The final images are presented
in Fig.~1.

\begin{figure*}
\centerline{\psfig{figure=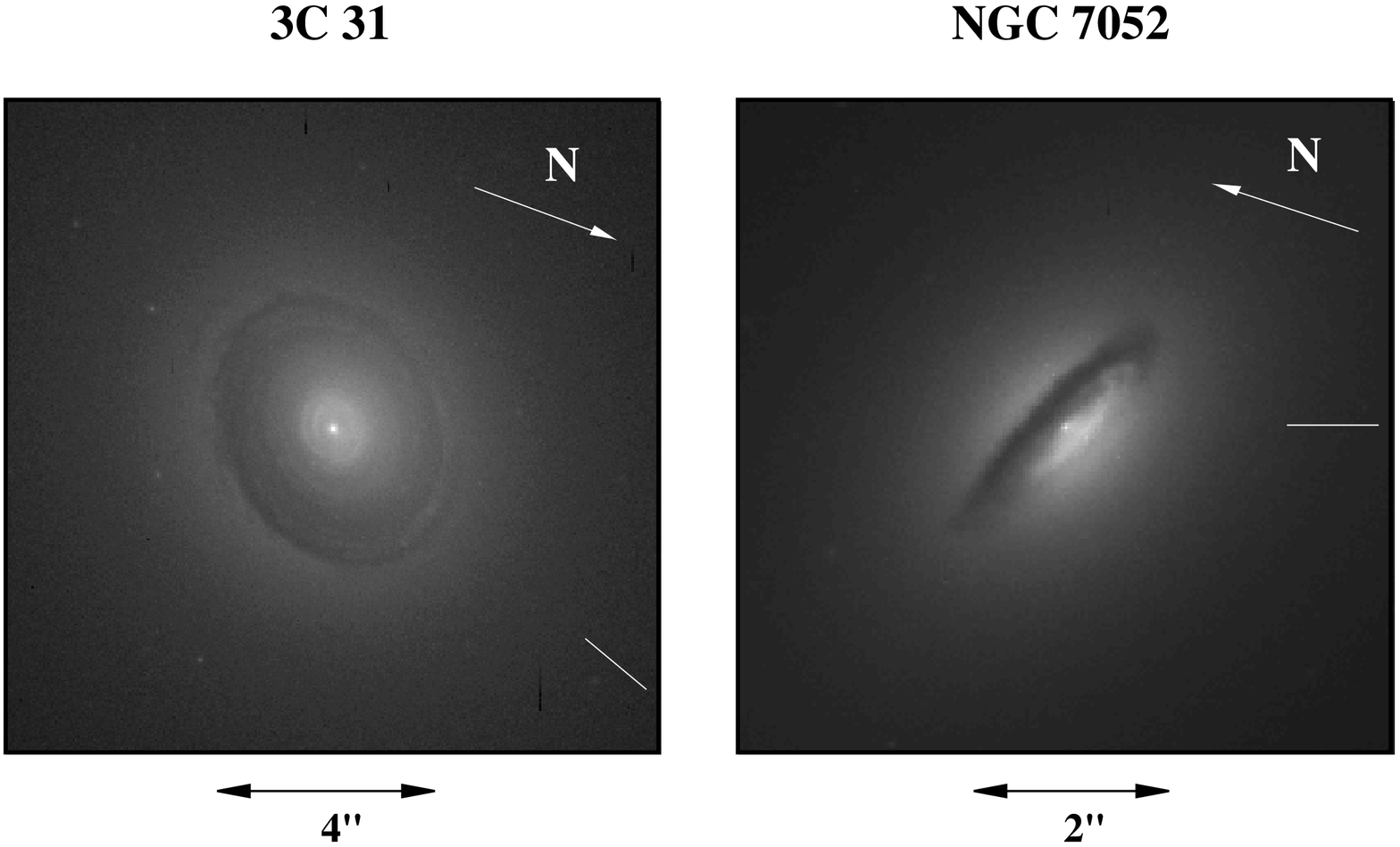,width=0.9\textwidth,angle=270}}
\end{figure*}

\begin{figure*}
\centerline{\psfig{figure=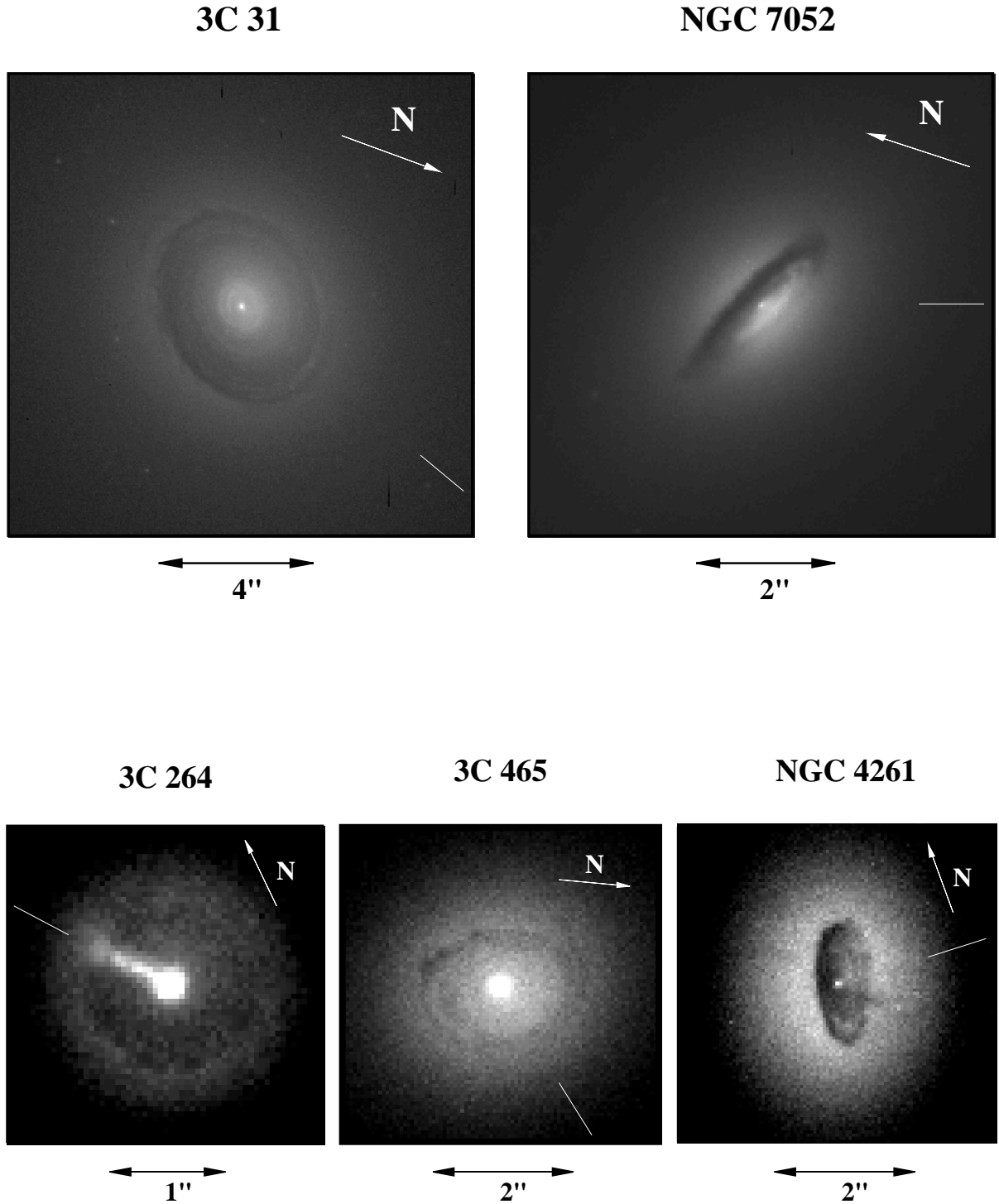,width=0.9\textwidth,angle=270}}
\caption{HST/WFPC2 broad band images of 3C~31, NGC~7052, 3C~264, 3C~465 and 
NGC~4261. The arrows indicate the North while the short dashes mark the orientation of the main radio jet.} 
\end{figure*}

\begin{table}
\caption{Log of HST observations for the five FR~I radio galaxies}
\begin{tabular}{|l c c c c |} \hline
Name &   Date   &    Filter  &   Range (\AA ) &   t$_{\rm exp}$ (s) \\
\hline

3C~264 &  24/12/94 &  F702W  & 6000--8000&   280\\
3C~31 &  21/01/95 &  F702W  & 6000--8000&   280\\
      &  01/09/95 &  FR680N & 6960--7050&   600\\
3C~465 &  23/01/95 &  F702W  & 6000--8000&   280\\
      &  23/08/95 &  F675N  & 6700--6770&   600\\
NGC~4261& 13/12/94 &  F547M  & 5200--5800&   800\\
NGC~7052& 23/06/95 &  F814W  & 7000--9000&  1400\\
\hline
\end{tabular}
\end{table}

At the center of each disc the HST images reveal the presence of a
compact nuclear source. In three cases (3C~264, 3C~31 and 3C~465) this
component is sufficiently bright, allowing us to derive its brightness
profile.  From this we measure FWHM ranging from 0$^{\prime\prime}$.06
to 0$^{\prime\prime}$.07, i.e. unresolved with the HST resolution at
7020 \AA. The upper limit to the physical size of these structures,
about one half of the resolution, is $\sim$ 10 -- 15 pc (cosmological
parameters $H_{\rm 0} = 75$ km s$^{-1}$ Mpc$^{-1}$ and $q_{\rm 0} =
0.5$ have been adopted throughout the paper).

Flux densities of these central sources are obtained with aperture
photometry and by adopting the internal WFPC2 flux calibration, which
is accurate to better than 5 per cent. The dominant photometric error
is the determination of the background in regions of varying
absorption and steep brightness gradients, particularly for the
faintest nuclei, which results in a typical uncertainty of 20 per
cent.

The narrow-band observations of 3C~31 and 3C~465 are used to remove
the line contamination from the most prominent emission lines included
in the spectral range covered by the broad-band filter, i.e. H$\alpha$
and [N II]. For 3C~31 the line emission contributes for 45 per cent to
the total measured flux density, while for 3C~465 it accounts for only
20 per cent. The luminosities corresponding to these corrected flux
densities are given in column 5 of Table~2.

Line contamination is not expected to be significant for NGC~4261,
observed with the F547M filter, which covers a relatively line-free
spectral region (5200 - 5800 \AA). However, this is not the case for
3C~264 and NGC~7052 which were observed in the spectral range 6000 -
8000 \AA \ and 7000 - 9000 \AA \ , respectively.

\begin{table}
\caption{FR~I radio galaxies}

\begin{tabular}{@{}lccccc} \hline
Name & $z$ & M$_{\rm R}$& Log P$_{\rm ext}$ & Log L$_{\rm o}$&  
Log L$_{\rm r}$  \\
(1) & (2) &(3) & (4) & (5)& (6)\\
\hline 
3C~264 &0.0216    &-22.70 &  25.21& 37.96 & 23.26 \\
3C~31 &0.0168    &-22.79 &  25.11& 36.78 & 22.89 \\
3C~465 &0.0291    &-23.32 &  25.90& 37.36 & 23.66 \\
NGC~4261 &0.0074  &-22.92 &  24.29& 35.34 & 22.28 \\
NGC~7052 &0.0164  &-22.10 &  23.59& 36.19 & 22.39 \\
\hline
\end{tabular}
\medskip 

Data for the five selected FR~I radio galaxies: (1) name, (2)
redshift, (3) galaxy R-magnitude, (4) extended radio power at 5 GHz in
W Hz$^{-1}$, (5) optical core luminosity in erg s$^{-1}$ \AA$^{-1}$ at
the wavelength indicated in Table 1, (6) radio core luminosity in erg
s$^{-1}$ Hz$^{-1}$ at 5 GHz.

\end{table}

\section{The relationship between discs and jets orientation} 

Let us now explore the geometrical relationship between disc and radio
axes and, in particular, whether and at which level of accuracy the
disc orientation can be considered an indicator of the jet direction.

In the simplest scenario radio jets are ejected along the rotation
axis of the central black hole, which also defines the orientation of
the inner accretion disc.  If the extended nuclear discs are coplanar
with the inner ones, we expect the radio jets to be perpendicular to
these structures.  When projected onto the plane of the sky, the radio
jets axis should therefore appear perpendicular to the (apparent)
major axis and parallel to the minor axis of the disc.

In order to test this hypothesis, we have derived, from high
resolution radio maps published in the literature (Morganti et
al. 1987; Morganti et al. 1993; Lara et al. 1997) the position angle
of the radio jets for the five radio galaxies. There are no
significant differences between the directions of pc and kpc jets
(e.g. Jones \& Wehrle 1997).  In Table~5 we report the position
angles relative to the brighter radio jet.

\begin{table*}
\caption{Jets and discs orientation}

\begin{tabular}{@{}lcccccccc}
\hline Name & Main Jet& Disc major& Disc incl. & Disc minor &
Jet/Disc & Minimum & Median & Median orientation\\ &P.A.($\gamma_{\rm J}$)
& axis ($\gamma_{\rm D}$) & ($\beta_{\rm D}$) & axis & offset & disc
warp & disc warp & difference ($\Delta \theta_{LOS}$)\\ \hline 
3C~264 & 27 & -100* & 15 & -10 & 37 & 9 & 31 (10-61) & 20 \\ 
3C~31 & -20 & -45 & 35 & 45 & -65 & 31 & 44 (32-63) & 24 \\ 
3C~465 & -54 & 10 & 45 & -80 & 26 & 17 & 24 (19-38) & 16 \\ 
NGC~4261 & -92 & -15 & 65 & -105 &  13 & 10 & 15 (12-25) & 10 \\ 
NGC~7052 & -165 & 65 & 72 & 155 & 40 & 36 & 39 (36-47) & 19 \\

\hline
\end{tabular}
\medskip 

{\small{* The major axis of 3C~264 has been determined 
as the separating line between the less and more obscured halves of 
the disc.}}

\end{table*}

The inclination of the extended nuclear discs with respect to the line
of sight is estimated by fitting ellipses to the sharp edges of these
absorption structures and with the reasonable assumption that they are
intrinsically circular. The typical uncertainty in the estimated disc
ellipticity is 5 per cent. In Table~5 we report the derived discs
inclinations $\beta_{\rm D }$ (i.e. the angle between the disc
symmetry axis and the line of sight) with relative errors.  From the
same fitting procedure we also derived the position angle of the disc
axis.

Before exploring in more detail the geometrical relationship between
discs and jets, it is useful to derive some additional information on
the geometry of the system. In all of our five cases, one half of the
disc produces a stronger obscuration than the other. This can be used
to identify which side of the disc itself is closer to us.
Furthermore, one radio jet is brighter than its opposite, an asymmetry
which is commonly ascribed to relativistic motion of the jet plasma at
kpc scale. Thus, for both the disc and the radio axes, we have no 180
degree position angle ambiguity.

Inspection of Table 5 shows that the main jet and disc axes are highly
correlated. This result, that the radio jet is always on the near side
of the dusty disc, is important since it adds even more weight to the
relativistic beaming explanation for the brightness and depolarization
asymmetries of radio jets (Garrington et al 1988; Morganti et al,
1997). It also suggests a causal connection between the dusty disc,
which may possibly have an origin external to the galaxy, and the
properties of the AGN. The radio jets, though reasonably well aligned
with the disc minor axes, are not exactly parallel. This implies that
either/both the overall AGN disc structure warps significantly from
the central to the extended accretion discs or/and that the observed
radio jets are not perpendicular the nuclear discs.

Two questions therefore arise at this stage: first of all, what is the
amplitude of the disc warps? and secondly, are the extended discs
useful indicators of the jet orientation? We will address these
questions in the next two sections.

\subsection{Discs warp angles}

If jets and discs orientations were completely unrelated we could
expect a random distribution of the differences in the projected
directions of main jet and disc minor axis, spanning between -180 and
180 degrees. Conversely, the measured offsets are found only between
-65 and +37. This indicates that discs and jets, although not
perpendicular, bear some geometrical relationship, which we will
quantify in the following.

It is possible to directly estimate the minimum intrinsic angle
$\Theta$ between the disc and the jet axis which can produce the
observed disc/jet offset. This, which might be interpreted as the
minimum amplitude of the disc warp from sub--pc to kpc scale, can be
expressed as
$$
\cos \Theta_{\rm min} = \cos \beta_{\rm J,min} \cos \beta_{\rm D} +
\cos(\gamma_{\rm J} - \gamma_{\rm D}) \sin \beta_{\rm J} \sin
\beta_{\rm D}
$$ 
where $\beta_{\rm J,min}$ is the orientation of the jet with respect to
the line of sight which minimizes $\Theta$ and is given by
$$ 
\tan \beta_{\rm J,min} = \cos(\gamma_{\rm J} - \gamma_{\rm D}) \tan
\beta_{\rm D}, 
$$
$\beta_{\rm J}$, $\beta_{\rm D}$, $\gamma_{\rm J}$ and $\gamma_{\rm
D}$ are the (observed/projected) angles defining the disc and jet
orientation (see Table~5).

Adopting the values reported in Table~5, the minimum warp angles range
between 9 and 36 degrees. Note that the warp angle can be smaller than
the observed projected offset. For example, in the case of 3C~31 an
offset of 65 degrees can be reproduced with a warp angle of only 30
degrees.

Furthermore, only a range of $\Theta$ can produce any observed
offset. As an example, in Fig.~2 we show the expected offset in the
case of 3C~465 ($\beta_{\rm D} = 45^\circ$) for different values of
the warp angles. An offset of 26$^\circ$ can be observed only for
$20^\circ \lesssim \Theta \lesssim 50^\circ$. More precisely, for each
object it is possible to estimate the probability of observing the
measured disc/jet offset for different values of the warp angle, as
shown in Fig.~3.  This probability is highly peaked at warp angles
close to their minimum value and then quickly decreases for larger
angles. The central values and the ranges of angles including 68\% of
these probability distributions are reported in Table~5.

It is important to notice that, from a simply geometrical point of
view, large warp angles are favored because of their statistical
weight (which is proportional to sin$\Theta$). Conversely, our
analysis points towards small values of $\Theta$ $\sim 20 - 40$
degrees, clearly indicating that discs and jets are not randomly
oriented one with respect to the other. The inclination of discs can
therefore be reasonably used to estimate the jet orientation. The
accuracy of this estimate is discussed in the next session.

\begin{figure}
\centerline{\psfig{figure=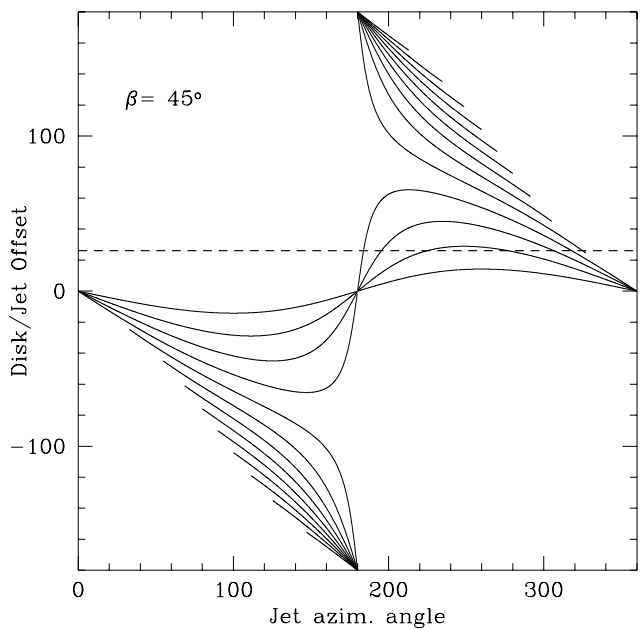,width=0.5\textwidth,angle=0}}
\caption{Projected offset between the main radio jet and the disc
minor axis, for warp angles increasing by 10 degrees steps from 10 to
130 degrees and for a disc inclination of 45 degrees, as in
3C~465. The dotted line represents the offset of 26$^{\circ}$ measured in
this radio galaxy. Note that offsets can be larger that the corresponding
warp angles and only a range of warp angles can correspond to a given
offset.}
\end{figure}

\begin{figure}
\centerline{\psfig{figure=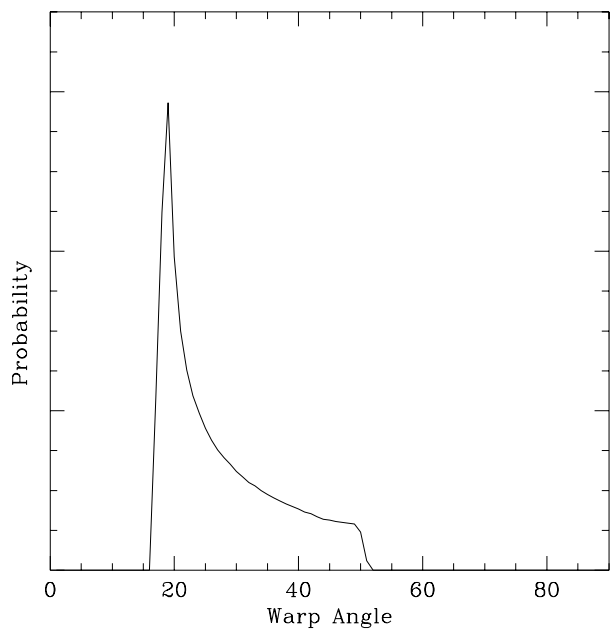,width=0.5\textwidth,angle=0}}
\caption{Probability (in arbitrary units) of observing an offset
within $\pm 2^\circ$ from the observed value for the case of 3C~465,
as a function of the warp angle $\Theta$. The probability is highly
peaked and its maximum is at an angle of $\sim 20^\circ$.}
\end{figure}

\subsection{Jet and disc orientation}

The angle which is relevant, in order to establish how good an
indicator of orientation the extended nuclear discs are, is the
difference in the orientations, $\Delta \theta_{LOS}$, of the jet axis
and the disc axis {\it with respect to the line of sight}.  We
estimated the median value of $\Delta \theta_{LOS}$ (reported in the
last column of Table 5) by adopting the probability distribution of
warp angles derived in the previous section.  We can conclude that the
disc orientation can be used of indicator of the jet orientation with
an uncertainty of $\sim$ 15 degrees. Furthermore, since there is no
bias favoring a particular range of orientation, this uncertainty can
be simply considered as a random error in the determination of the
radio axis.

\section{Comparison between radio galaxies and BL Lacs} 

\subsection{The selection of the beamed counterparts}

Within the considered unification model, FR~I and BL Lacs are
essentially a single class of objects only viewed at different
orientations.  But, when comparing randomly selected pairs of objects,
a large scatter in their properties is obviously expected. However,
correlations between the core radio power and both the extended radio
and host galaxy luminosities have been established (Giovannini et
al. 1988). This implies than when we select FR~I and BL Lacs with
identical isotropic quantities, their beamed properties will be
distributed within approximately the scatter in the above correlation.

Data relative to the isotropic properties for the five FR~I radio
galaxies are reported in Table~2. Absolute R magnitudes (aperture
corrected, Sandage 1972), are derived from Sandage (1973) and Impey \&
Gregorini (1993). Radio luminosities are estimated from flux densities
given by Impey \& Gregorini (1993) and K\"uhr et al. (1981).

We then select a sample of BL Lacs for which both extended radio and
galaxy optical flux densities are reported in the literature. The data
are summarized in Table~3. Magnitudes of the BL Lacs host galaxies are
taken from Wurtz et al. (1996). They have been scaled according to our
assumptions on the cosmological parameters and Pence's (1976)
K-correction has been applied.\footnote{Given the level of accuracy
reached in the comparison of BL Lacs and FR~I properties, we can
neglect here the small correction between the r-Gunn (used by Wurtz et
al. 1996) and the standard R Johnson filters.} The galaxy magnitudes
span between $-21$ and $-24$. Luminosities of the extended radio
emission at 5 GHz were taken from Perlman \& Stocke (1993) and have
been similarly corrected, adopting an energy spectral index of
0.7. They cover the range $10^{22.5}-10^{27}$ W Hz$^{-1}$.

In order to select the BL Lacs objects to compare with each of the
FR~I galaxies, we plot all the sources (FR~I and BL Lacs) in the host
galaxy R--magnitude (M$_{\rm R}$) vs extended radio power at 5 GHz
(P$_{\rm ext}$) plane (Fig.~4). Here the FR~I radio galaxies are all
located in the portion of the M$_{\rm R}$ -- P$_{\rm ext}$ plane
covered by BL Lac objects. The BL Lac counterparts of each radio
galaxy have been then chosen among those which differ by less than a
factor of two in both radio and optical luminosities.  For 3C~264 and
3C~465, which lie in a relatively more sparsely populated region of
the plane, we allow for the slightly larger difference in M$_R$ of 1
mag.  We thus identified between 2 and 5 beamed counterpart for each
`parent' radio galaxy. The complete list of all the `relative' BL Lacs
(with their optical and radio core properties) is reported in Table~3.

\begin{figure}
\centerline{\psfig{figure=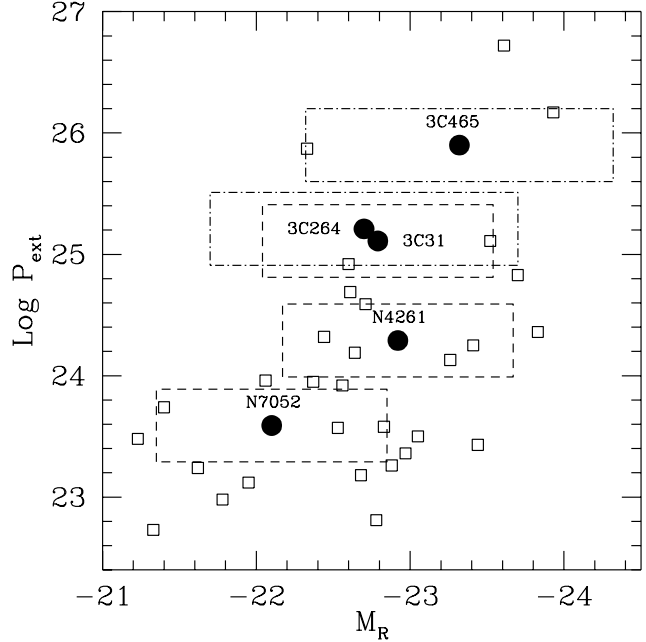,width=0.5\textwidth}}
\caption {Extended radio power at 5 GHz versus host galaxy absolute R
magnitude for BL Lacs and FR~I radio galaxies. BL Lacs included in the
boxes differ by less than a factor of two in both radio and optical
luminosities and have thus been selected as counterparts of our FR~I.}
\end{figure}

\begin{table}
\caption{BL Lac objects}

\begin{tabular}{@{}lccccc}\hline
Name & $z$ & M$_{\rm R}$&  Log P$_{\rm ext}$ &Log L$_{\rm o}$& 
Log L$_{\rm r}$ \\
(1) & (2) &(3) & (4) & (5)& (6)\\
\hline
0548--322 (R)&0.069 &-22.44  &24.32 & 40.33 & 23.90 \\
0851+202 (R)&0.306  &-21.40  &23.74 & 41.83 & 26.84 \\ 
1400+162 (R)&0.244  &-22.33  &25.87 & 40.98 & 25.56 \\
1407+599 (X)&0.495  &-23.52  &25.11 & 40.43 & 25.08 \\
1443+638 (X)&0.299  &-22.64  &24.19 & 40.01 & 24.32 \\
1458+228 (X)&0.235  &-22.53  &23.57 & 40.91 & 24.63 \\ 
1538+149 (R)&0.605  &-23.93  &26.17 & 41.38 & 27.27 \\
1552+203 (X)&0.222  &-23.41  &24.25 & 40.50 & 24.62 \\
2007+777 (R)&0.342  &-22.60  &24.92 & 41.33 & 26.53 \\
2143+070 (X)&0.237  &-22.71  &24.59 & 40.43 & 24.82 \\
2200+420 (R)&0.069  &-22.83  &23.58 & 40.57 & 25.68 \\
2254+074 (R)&0.190  &-23.26  &24.13 & 40.87 & 25.61 \\
\hline
\end{tabular}
\medskip 

Data relative to BL Lacs: (1) name and selection band (R=radio
selected, X=X--ray selected, from Perlman \& Stocke 1993), (2)
redshift, (3) galaxy magnitude R, (4) extended radio power at 5 GHz in
W Hz$^{-1}$, (5) optical core luminosity in erg s$^{-1}$ \AA$^{-1}$ at
5000 \AA, (6) radio core luminosity in erg s$^{-1}$ Hz$^{-1}$ at 5
GHz.
\end{table}

\subsection{Comparison of the core properties}

The core optical luminosities for the radio galaxies are estimated
from HST observations (see Section 2), while the radio core
luminosities are from VLBI observations by Giovannini, Feretti \&
Comoretto (1990) and Jones et al.  (1981). Radio and optical core
emission of the selected BL Lac objects were taken from Giommi \&
Padovani (1993).  For BL Lacs, energy spectral indices $\alpha=0$ in
the radio and $\alpha=1$ in the optical bands were adopted to estimate
the K--correction ($F(\nu)\propto \nu^{-\alpha}$).

The list of the FR~I--BL Lac pairs, together with the results of the
comparisons for each of the association FR~I--BL Lac, are presented in
Table~4. In particular, in columns 3 and 4 the ratios of the parent
FR~I and BL Lac core luminosities $R=L_{\rm BL Lac}/L_{\rm FR I}$, are
reported, for the optical and radio bands, respectively.  These
measured ratios (which can be as large as $3 \times 10^5$) clearly
indicate that any optical non-thermal emission in FR~I radio galaxies
is dramatically fainter than in BL Lacs with similar extended
properties. There appears to be no relation between the estimated $R$
and the spectral type/selection band of the considered BL Lac objects.

\begin{table}
\caption{Results: FR~I vs BL Lacs nuclear luminosity ratios}
\begin{tabular}{@{}lcccccccc}\hline
Name & Relatives & R$_{\rm o}$ & R$_{\rm r}$ \\
(1) & (2) &(3) & (4) \\
\hline

3C~264   &1407+599 & 2.23e2 & 6.61e1 \\
         &2007+777 & 1.77e3 & 1.86e3 \\
3C~31    &1407+599 & 3.54e3 & 1.55e2 \\ 
         &2007+777 & 2.81e4 & 4.37e3 \\ 
3C~465   &1400+162 & 1.12e3 & 7.94e1 \\
         &1538+149 & 1.12e4 & 4.07e3 \\
NGC~4261 &0548--322& 9.27e4 & 4.17e1 \\
         &1443+638&  4.64e4& 1.10e2 \\
	 &1552+203& 1.47e5 & 2.19e2 \\
	 &2143+070& 1.17e5 & 3.47e2 \\
	 &2254+074& 2.92e5 & 2.14e3 \\
NGC~7052 &0851+202& 2.99e5 &2.82e4 \\
        &1458+228&  3.59e4 &1.74e2  \\ 
        &2200+420&  1.64e4 &1.95e3 \\ 
\hline
\end{tabular}
\medskip 

Summary of the results, for each FR~I -- BL Lac pair: (1) Radio
galaxy; (2) associated BL Lacs; (3) BL Lac vs FR~I core luminosity
ratio in the optical band; (4) core radio luminosity ratio.

\end{table}

Furthermore, we stress that these are only lower limits to the actual
ratios of BL Lac/FR~I jet luminosities, since an unknown fraction of
the optical emission associated to the central regions of radio
galaxies can be either stellar in origin and/or might be produced by
the innermost regions of an accretion disc.

The ratios $R_{\rm r}$ at radio wavelengths are significantly smaller
(with typical values $\sim 10^2$ -- 10$^4$) than those measured in the
optical band, and also in this case they must be considered lower
limits as there might be radio emission not associated with the
relativistic jet.

>From Table~4 we can also see that the spread of the optical ratios for
a given FR I radio galaxy is about an order of magnitude. This scatter
includes and is larger than the dispersion in the correlation between
core and extended radio-luminosity in radio--galaxies, which shows a
logarithmic scatter of only $\sim 0.7$ (Giovannini et al. 1988).

\section{Evidence for beamed emission in FR~I}

On the basis of the results of the Section 3, it becomes
meaningful to examine whether there is any dependence of the estimated
luminosity ratios, $R_{\rm o}$ and $R_{\rm r}$, on the orientation of
the radio galaxy. Either the presence or absence of a trend would be
an important clue for the understanding of the amount of beamed vs
isotropic radiation fields in radio galaxies.

In Fig.~5 we show the ratio $R_{\rm o}$ between the optical core
luminosity of the five FR I radio galaxies and the corresponding BL
Lacs as a function of the radio galaxy inclination angle.  The
vertical bar for each radio galaxy has been obtained by considering
all the BL Lacs--FR~I pairs, as indication of the scatter.  The
horizontal bars represent instead the median disc warp angles, to take
into account the uncertainty on the relative orientation of the jet
axis and the line of sight discussed in Section 4.  Similarly, we
estimated the ratio between the core luminosity of radio galaxies and
BL Lacs at radio wavelengths, $R_{\rm r}$ (see Fig. 6 and Table 4).

Figure 5 shows that there is a large increase of the optical ratio
$R_o$ with the inclination of the nuclear discs. The trend is
qualitatively in the sense expected if the emission from the central
nuclei observed in the radio galaxies is the less beamed counterpart
of the jet emission observed in BL Lacs. Unfortunately, with only 5
objects, the correlation is not formally statistically significant and
the results are only suggestive. In addition to beaming, other
explanations of the trend seen in Fig 5. are possible and these are
discussed in the next section.

\begin{figure}
\centerline{\psfig{figure=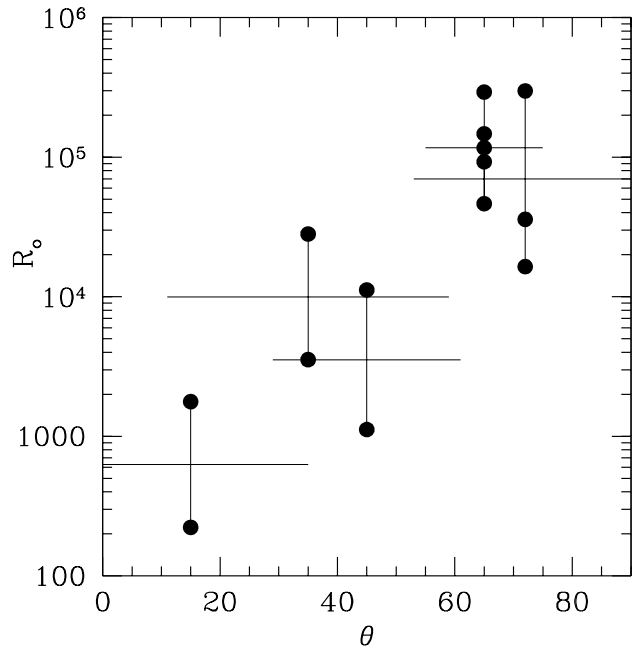,width=0.5\textwidth,angle=0}}
\caption{Ratios between the optical core luminosity of each of the
five FR~I radio galaxies with the corresponding BL Lacs, as a function
of the orientation of the nuclear disc. Note the large increase of
the ratio with increasing inclination of the radio galaxy axis with
respect to the line of sight.}
\end{figure}

\begin{figure}
\centerline{\psfig{figure=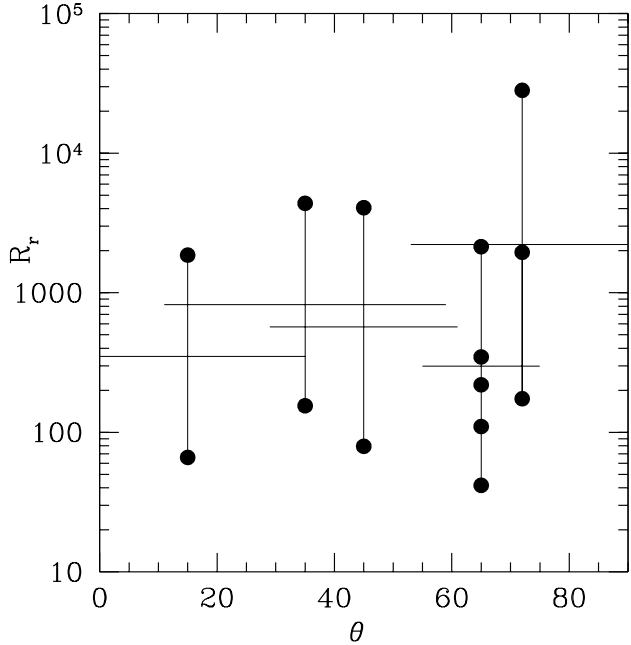,width=0.5\textwidth,angle=0}}
\caption{Ratios between the radio core luminosity of each of the
five FR~I radio galaxies with the corresponding BL Lacs, as a function
of the orientation of the nuclear disc.}
\end{figure}

\subsection{Alternative explanations for a $R_{\rm o}$ vs $\theta$
dependence}

There are at least two other possible interpretations of a dependence of
$R_o$ on $\theta$ other than beamed synchrotron emission: we might be
observing the anisotropic emission from a thick accretion disc;
alternatively, a central source, such as a stellar cusp, might be
obscured by an optically and geometrically thick absorbing structure
whose optical depth varies with $\theta$. In both cases the
correlation between $R_{\rm o}$ and $\theta$ would be due to a
dependence of the {\it observed} core luminosity $L_{\rm core}$ on
inclination which is not to ascribe to beaming.

Let us consider these alternative explanations in turn.  It has been
shown that thick accretion discs might produce a significantly
anisotropic radiation field (e.g. Sikora 1981, Madau 1988). However,
although this anisotropy can be large at UV and X-ray wavelengths, a
face-on disc would appear only a few times brighter than an edge-on
disc in the optical band. Given the much larger dependence on
inclination we observe this effect does not appear to dominate in this
context.

We now examine the expected trend of $L_{\rm core} (\theta)$ to
explore the possible role of absorption.  At least two cases are
viable and the central source might be: i) of fixed luminosity; ii) a
fixed fraction of the total starlight, $L_{\rm gal}$. Almost
independently of the adopted normalization a trend of luminosity with
inclination is present (see Fig.~7) with an decrease in $L_{\rm core}$
of approximately two orders of magnitude from low to high values of
$\theta$.

If absorption is the main cause of this behavior, the optical depth of
any obscuring material must increase with $\theta$.  Adopting a
standard gas-to-dust ratio, for which the extinction is related to the
neutral hydrogen column density by $A_{\rm V} \sim 5 \times 10^{-22}
N_{\rm H}$, the trend of luminosity with inclination is reproduced if
$N_{\rm H}$ increases, approximately linearly, by $\Delta N_{\rm H}
\sim 10^{22}$ cm$^{-2}$ from $\theta = 15^{\circ}$ to $\theta =
72^{\circ}$.  While this value can be quantitatively plausible, it is
very tightly constrained: an increase in $\Delta N_{\rm H}$ by only an
extra factor of 3 would lead to a much stronger dependence of $L_{\rm
core} (\theta)$, with a change by six orders of magnitudes between the
low and high inclination galaxies. On the other hand, a decrease of
the same factor in $\Delta N_{\rm H}$ would not produce a significant
change of $L_{\rm core}$ with $\theta$. We think that the required
quite well defined dependence of the absorbing material optical depth
on its geometrical structure {\it in different objects} which are also
of different radio and optical luminosities, makes this interpretation
unappealing.

We conclude that both of these alternatives appear implausible and
that the observed anisotropy is most likely produced by the effect of
Doppler beaming.

\begin{figure}
\centerline{\psfig{figure=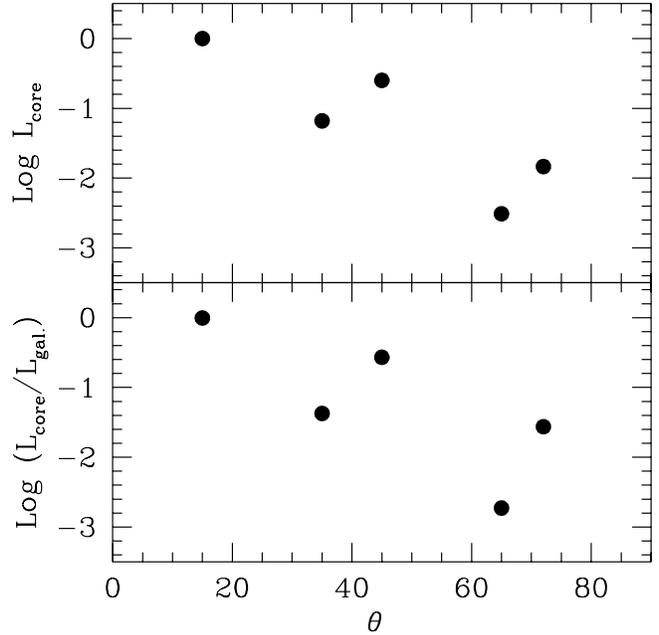,width=0.5\textwidth,angle=0}}
\caption{Optical luminosity of the radio galaxies cores $L_{\rm core}$
versus orientation $\theta$, normalized to the value of 3C~264 (top
panel); ratio of $L_{\rm core}$ and galaxy luminosity, again
normalized to 3C~264 (bottom panel)}
\end{figure}

\section{Discussion}

\begin{table}
\caption{Results: FR~I inclination angles, luminosity ratios and
derived Lorentz factors}
\begin{tabular}{@{}lccccc}\hline
Name & Relatives & $\theta_{\rm FRI}$ & $\Gamma_{\rm o, min}$ 
& $\Gamma_{\rm r, min}$ & Error \\
(1) & (2) &(3) & (4) & (5) & (6) \\
\hline
3C~264   &1407+599& 15 &  8.7 (6.5) & 10.3 (6.7) & 40 \% \\
         &2007+777&    & 12.8 (9.0)    & 24.9 (12.9) & \\
3C~31    &1407+599& 35 & 6.3(4.4) & 5.7(3.6) & 30 \% \\ 
        &2007+777&     & 9.1 (5.8) & 13.4 (6.6) &\\ 
3C~465   &1400+162& 45 &   4.1  (3.0) & 3.8 (2.8) & 25 \% \\ 
        &1538+149&     &   6.1  (4.1) & 10.4 (5.1) &\\ 
NGC~4261 &0548--322& 65 & 6.5 (4.0) & 2.3 (1.7) & 15 \% \\
        &1443+638& &   5.8  (3.6) & 3.0 (2.0) &\\ 
        &1552+203& &   7.0  (4.2) & 3.6 (2.3)&\\
        &2143+070& &   6.7  (4.1) & 4.0 (2.4) &\\ 
        &2254+074& &  7.8  (4.6) & 6.3 (3.3) &\\
NGC~7052 &0851+202& 72   &  7.0  (4.1) & 10.6 (4.5) &15 \% \\ 
        &1458+228& &   4.9  (3.2) & 3.0 (2.0) &\\ 
        &2200+420& &  4.3  (2.9) & 5.5 (2.9) & \\ 
\hline

\end{tabular}

\medskip 

Summary of the results, for each FR~I -- BL Lac pair: (1) Radio
galaxy; (2) associated BL Lacs; (3) radio galaxy orientation, as
inferred from the nuclear disc with respect to the line of sight (in
degrees); estimated $\Gamma_{\rm min}$
in the optical (4) and radio (5) bands; (6) relative error on $\Gamma$
deriving from uncertainties on $\theta_{\rm FRI}$. The values of
Lorentz factors reported in the Table refer to the cases $p=2$ and
$p=3$, the latter shown in parenthesis (see Section 6.1 for the
definition of $p$).

\end{table}

It is now possible to compare, at least at the zero order, the
quantitative predictions of the beaming model concerning the
dependence of luminosity on the inclination angle. In the simplest
scenario, the beamed emission originates from a relativistic jet
described by a single Lorentz factor $\Gamma$, which, by hypothesis,
is the same for BL Lac and FR~I.  In this case the ratio $R$ between a
BL Lacs and its parent radio galaxy luminosity is given by
$$ R = ((1 - \beta\ \cos \theta_{\rm FRI}) / (1 - \beta\ \cos
\theta_{\rm BL Lac}))^{(p+\alpha)},
$$ 
where the angles $\theta$ are measured with respect to the beaming
(jet) axis, $\alpha$ is the spectral index in the considered energy
band and $\beta c$ the plasma velocity. The exponent $p$ depends on
the geometrical and temporal structure of the emitting plasma,
e.g. whether the emission region can be better approximated as a
continuous jet or an individual 'blob' of plasma; plausibly in a
realistic situation $ 2 \lesssim p \lesssim 3$.

The comparison with the observed values of $R$ reported in Table~4,
therefore allows to estimate the {\it minimum} jet Lorentz factors
$\Gamma_{\rm min}$ consistent with dimming the jet emission (assumed
to be produced by a BL Lac observed along the jet axis) by the
observed ratio when moving from $\theta=0$ to $\theta_{\rm FRI}$.  The
resulting lower limits on the jet Lorentz factor, for each FR~I--BL
Lac pair, in the two energy bands and for different values of $p$ are
reported in Table~5 and shown in Fig.~8.

\begin{figure}
\centerline{\psfig{figure=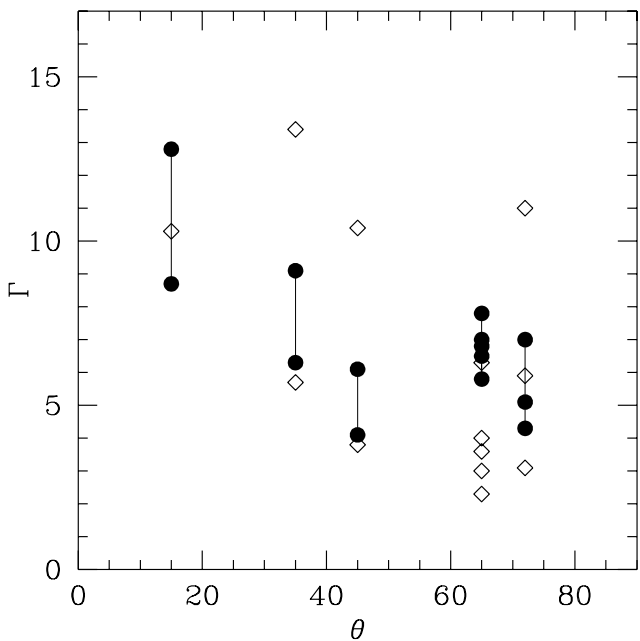,width=0.5\textwidth,angle=0}}
\caption{Lower limits to the jet Lorentz factors for each FR~I--BL Lac
pair, derived from the luminosity ratio of the nuclear sources. Limits
are estimated in both the optical (filled circles) and radio bands (diamonds) for $p=2$ (see text). 
For clarity of the figure, the Lorentz factor
$\Gamma=24.9$ in the radio case for 3C~264 has not been
plotted}
\end{figure}

The formal estimate of a lower limit for $\Gamma$ corresponds to
$\theta_{\rm BL Lac} = 0$. However, it is unlikely that all BL Lacs
jets are so closely aligned with the line of sight.  We therefore
self-consistently estimated a less conservative $\Gamma^*$,
considering $\theta_{\rm BL Lac}$ as the average angle within the
1/$\Gamma^*$ beaming cone. The resulting $\Gamma^*$ are 15 - 25 per
cent larger than $\Gamma_{\rm min}$.

Several sources of error are involved in the calculation of
$\Gamma_{\rm min}$, but they are all dominated by the uncertainties in
the jet orientation and by any intrinsic scatter between isotropic and
beamed properties. The former ones translate into an error of $\sim 5
- 30$ per cent on $\Gamma_{\rm min}$, increasing for smaller
$\theta_{\rm FR I}$, while the values of $\Gamma_{\rm min}$ obtained
by using different associated BL Lacs for the same radio galaxy are
spread by less than $\pm$ 25 per cent. Photometric errors are
negligible in view both of the variability of BL Lacs and the weak
dependence of $\Gamma$ on the derived ratios $R$.

In Fig.~9 we plot the expected $R_{\rm o}(\theta)$ curves for
different values of $\Gamma$, within the estimated range. Clearly,
these curves completely bracket the measured luminosity ratios, and
follow the general trend.

The dependence of $R_{\rm r}$ (i.e. the luminosity ratio in the radio
band) on $\theta$, presented in Fig. 6, is considerably weaker than in
the optical band with an overall increase of about one order of
magnitude. This might be in part expected in the beaming scenario,
since the lower spectral index of radio ($\alpha = 0$) with respect to
optical ($\alpha = 1$) emission makes the anisotropy less pronounced
at these wavelengths.  However, the limits on $\Gamma$ derived from
the radio ratios are overall comparable with, although with a larger
spread than, the optical ones.

The main point we would like to stress is that the derived Lorentz
factors, both in the optical and radio bands, are completely
consistent with those derived by other means (as superluminal
velocities, synchrotron self--Compton model, core dominance
parameters, number densities estimates, e.g. Ghisellini et al. 1993).
Note that the main constraint on the values of $\Gamma$ derived above
is imposed by the large range spanned by $R(\theta)$, independently of
the particular shape of the $R(\theta)$ dependence.

We conclude that our results can be well accounted for if the optical
nuclear sources seen in these FR~I radio-galaxies are indeed the
un-beamed counterpart of BL Lac objects, in agreement with the basic
features of the unifying scheme for these two classes of objects.
 
\begin{figure}
\centerline{\psfig{figure=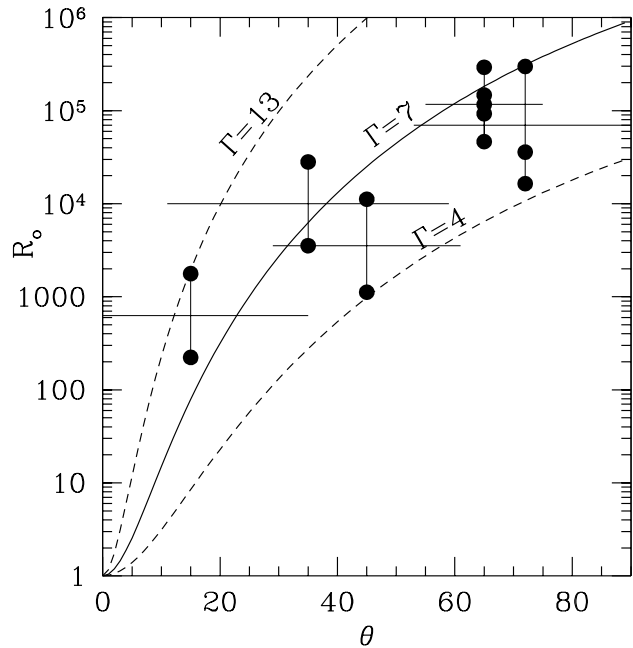,width=0.5\textwidth,angle=0}}
\caption{ $R_{\rm o}({\theta})$ curves for different values of the jet
Lorentz factors $\Gamma$ estimated for $p$=2.}
\end{figure}

\section{Summary and future prospects} 

Two are the main results of this paper.  We examined the relative
orientation of the observed nuclear dusty disc structures observed
with HST for five radio galaxies with that of their (VLA) radio
jets. We find that, though these two axis are not parallel, they are
clearly related. The most probable intrinsic disalignement/warp is
between 20 and 40 degrees. Interestingly, this quite consistent with
theoretical expectations (Natarajan \& Pringle 1998).  However, this
misalignment translates into an angular difference between the
directions of the line of sight with respect to the jet and disc axis
of only 10 - 20$^\circ$.  The nuclear discs are therefore reasonable
indicators of the symmetry axis of the system.

The second finding we would like to emphasize is that the optical
radio-galaxies cores are between $2 \times 10^2$ and $3 \times 10^5$
times fainter than those seen in BL Lac objects with matching extended
properties and that this ratio shows a suggestive dependence on the
radio galaxy orientation. Although extinction might reasonably play a
role, we think it is rather implausible that this is the main cause of
this trend. Conversely, this dependence can then be ascribed to the
presence of an anisotropic nuclear component in radio galaxies, the
counterpart of the beamed emission which dominates in BL Lac objects.
Indeed, these results can be successfully accounted for if we are
seeing emission from a relativistic jet with a Lorentz factor $\Gamma
\sim 5 - 10$. These values are consistent with those derived from both
the ratios of the nuclear radio luminosities and the quantitative
predictions of the BL Lac/ FR~I unified model.

Several observational tests can be performed to confirm and strengthen
our findings.

First of all, the prevalence of non-thermal over stellar emission in the
central regions of FR~I can be confirmed through a spectral analysis,
e.g. by measuring the amplitude of the 4000 \AA\ calcium II break.
Moreover, the central sources seen in the FR~I are expected to show
the typical behavior of the BL Lac cores. i.e. large polarization and fast
variability.

Clearly, information on a much larger sample of FR~I radio sources
will be very helpful to improve the statistical significance of our results since many more FR~I--BL Lac pairs can then be formed. It will also be possible to compare the nuclear luminosity of radio galaxies
with similar extended properties but with different orientation.
This will allow a much better determination of the $R(\theta)$ curve 
and provide further clues on the FR~I and BL Lacs relationship.
In particular the luminosity ratio dependence on angle can set upper
limits to any isotropic/non--beamed emission, a critical and unknown
parameter when comparing the luminosity functions of FR~I and BL Lacs
(Padovani \& Urry 1991; Urry, Padovani \& Stickel 1991; Celotti et al. 1993).  In fact, the presence of any isotropic emission would cause $R$ to saturate to a constant when this component becomes dominant at large $\theta$.  
Furthermore, the $R(\theta)$ dependence would constraint the height of any  obscuring torus, since it will enable us to establish the largest value of $\theta$ for which we have a direct view of the FR~I nuclei.
Finally, the comparison with a large enough (and complete) sample of
FR~I can be performed statistically even without indication on the
orientation of the individual objects. The distribution of the
observed ratios, rather than individual values, would then be compared
with that expected from a random orientation of radio galaxies in the
sky. This improvement in the sample of FR~I is expected to
be produced by the snapshot survey of B2 radio galaxies currently under
execution with HST.
  
\section*{Acknowledgments} 
We thank the referee, Ian Browne, for his criticisms and suggestions,
which greatly helped improving the paper. Both ACs acknowledge the Italian
MURST for financial support.

\section*{References} 

\refitem Abraham R.G., McHardy I.M., Crawford C.S., 1991, MNRAS, 252,
482

\refitem Antonucci R.R.J., Ulvestad J.S., 1985, ApJ, 294, 158

\refitem Bailey J., Sparks W.B., Hough J.H., Axon D.J., 1986, Nat,
322, 150

\refitem Blandford R.D., Rees M.J., 1978, in BL Lac Objects,
A.N. Wolfe, ed., Univ. Pitt. Press (Pittsburgh), p.~328

\refitem Burrows C.J., et al., 1995, WFPC2 Instrument Handbook,
Version 3.0, Space Telescope Science Institute

\refitem Celotti A., Maraschi L., Ghisellini G., Caccianiga A.,
Maccacaro T., 1993, ApJ, 416, 118

\refitem De Juan L., Colina L., Golombek D., 1996, A\&A, 305 776

\refitem De Koff S., Baum S., Sparks W.B., Biretta J., Golombek D.,
Macchetto F.D., McCarthy P.J., Miley G.K., 1995, BAAS, 187, 1202

\refitem Falomo R., Urry C.M., Pesce J.E., Scarpa R., Giavalisco M.,
Treves A., 1997, ApJ, 476, 113

\refitem Fanaroff B.L., Riley J.M., 1974, MNRAS, 167, 31p

\refitem Garrington S.T., Leahy J.P., Conway R.G., Laing R.A., 1988,
Nat, 331, 147

\refitem Ghisellini G., Maraschi L., 1989, ApJ, 340, 181

\refitem Ghisellini G., Padovani P., Celotti A., Maraschi L., 1993,
ApJ, 407, 65

\refitem Giommi P., Padovani P., 1993, MNRAS, 277, 1477

\refitem Giovannini G., Feretti L., Comoretto G., 1990, ApJ, 358, 159

\refitem Giovannini G., Feretti L., Gregorini L., Parma P., 1988,
A\&A, 199, 73

\refitem Impey C., Gregorini L., 1993, AJ, 105, 853

\refitem Jaffe W., Ford H.C., Ferrarese L., Van den Bosch F.,
O'Connell R.,W.  1993, Nat, 364, 213

\refitem Jones D.L., Sramek R.A., Terzian Y., 1981, ApJ, 246, 28

\refitem Jones D.L., Wehrle, A.E., 1997, ApJ, 484, 186

\refitem Kollgaard R.I., Wardle J.F.C., Roberts D.H., Gabuzda D.C.,
1992, AJ, 104, 1687

\refitem K\"uhr H., Witzel A., Pauliny-Toth I.I.K, Nauber U., 1981,
A\&AS, 45, 367

\refitem Lara, L., Cotton W.D., Feretti, L., Giovannini, G., Venturi,
T., Marcaide, J.M., 1997, ApJ, 474, L179

\refitem Madau P., 1988, ApJ, 327, 116

\refitem Morganti R., Fanti, C., Fanti, R., Parma, P., de Ruiter, H.R.,
1987, A\&A, 183, 203

\refitem Morganti R., Fosbury R.A.E., Hook R.N., Robinson A.,
Tsvetanov Z., 1992, MNRAS, 256, 1

\refitem Morganti R., Robinson A., Fosbury R.A.E., di Serego Alighieri
S., Tadhunter C.N., Malin D.F., Hook R.N., 1991, MNRAS, 249, 91

\refitem Morganti, R., Killeen, N.E.B., Tadhunter, C.N., 
1993, MNRAS, 263, 1023 

\refitem Morganti R., Parma, P., Capetti, A., Fanti, R., de Ruiter, H.R.,
1997, A\&A 326, 919

\refitem Murphy D.W., Browne I.W., Perley R.A., 1993, MNRAS, 264, 298

\refitem Natarajan P., Pringle, J.E., 1998, ApJL, in press
	
\refitem Packham C., Hough J.H., Efstathiou A., Chrysostomou A.,
Bailey J.A., Axon D.J., Ward M.J., 1996, MNRAS, 278, 406

\refitem Padovani P., Urry C.M., 1990, ApJ, 356, 75

\refitem Padovani P., Urry C.M., 1991, ApJ, 368, 373

\refitem Pence W., 1976, ApJ, 203, 39

\refitem Perlman E.S. , Stocke J.T., 1993, ApJ, 406, 430

\refitem Sandage A., 1972, ApJ, 178, 1

\refitem Sandage A., 1973, ApJ, 183, 711

\refitem Schreier E.J., Capetti A., Macchetto F.D., Sparks W.B., Ford
H.J., 1996, ApJ, 459, 535

\refitem Sikora M., 1981, MNRAS, 196, 257

\refitem Stickel M., K\"uhr H., 1993, A\&A, 98, 39

\refitem Stickel M., Padovani P., Urry C.M., Fried J.W., K\"uhr H.,
1991, ApJ, 374, 431

\refitem Sutherland, R.S., Bicknell, G.V., Dopita, M.A., 1993, ApJ,
414, 506

\refitem Ulrich M-H, 1989, in BL Lac Objects: 10 years after, Maraschi
L., Maccacaro T. \& Ulrich M.-H. eds, Springer-Verlag, p. 45

\refitem Urry C.M., Padovani P., 1995, PASP, 107, 803

\refitem Urry C.M., Padovani P., Stickel M., 1991, ApJ, 382, 501

\refitem Wurtz R., Stocke J.T., Yee H.K.C., 1996, ApJS, 103, 109

\end{document}